\documentclass[showpacs,preprintnumbers,prd,nofootinbib,floats,amssymb,floatfix]{revtex4}
\usepackage{amsmath}
\usepackage{graphicx}
\usepackage{amsxtra}
\usepackage{hyperref}
\usepackage{amsmath}
\usepackage{amstext}
\begin{document}
\title{ Gauge symmetry and constraints structure in topologically massive AdS gravity: A symplectic viewpoint}
\author{ Omar Rodr{\'i}guez-Tzompantzi}  \email{omar.tz2701@gmail.com}
 \affiliation{ Facultad de Ciencias F\'{\i}sico Matem\'{a}ticas, Benem\'erita Universidad Au\-t\'o\-no\-ma de Puebla,
 Apartado postal 1152, Puebla, Pue., M\'exico}
\author{Alberto Escalante}  \email{aescalan@ifuap.buap.mx}
 \affiliation{  Instituto de F{\'i}sica, Benem\'erita Universidad Aut\'onoma de Puebla, \\
 Apartado Postal J-48 72570, Puebla Pue., M\'exico, }
\begin{abstract}
By applying the Faddeev-Jackiw symplectic approach we systematically show that both  the local gauge symmetry  and the constraint   structure  of topologically massive  gravity with a cosmological constant $\Lambda$, elegantly encoded in the  zero-modes of the symplectic matrix, can be identified. Thereafter, via an  appropriate partial gauge-fixing procedure, the time gauge, we calculate the quantization bracket structure (generalized Faddeev-Jackiw brackets) for the dynamic variables and confirm that the number of physical degrees of freedom is one. This approach provides an alternative to explore the dynamical content of massive gravity models.
\end{abstract}
 \date{\today}
\preprint{}
\maketitle
\section{Introduction}
Fundamental issues in modern cosmology, such as inflation, dark matter and dark energy \cite{ac-uni,dark}, which attempt to explain the primordial and late time accelerating expansion of our universe, have long been  motivated alternative gravity theories beyond original Einstein's General Relativity, both in the ultraviolet (UV)  and the infrared (IR) regimes. According to  Lovelock's theorem \cite{lov1,lov2} any modification of General Relativity requires at least one of the following ingredients:  i) extra dimensions, ii) extra degrees of freedom, iii) higher-derivatives terms, and iv) non-locality.  Massive gravity theories are an example of  the type-ii ingredients, in which the massless graviton of General Relativity is given a non-zero mass    (see e.g. \cite{Infraded, Rham,Rham1,Rham2,Rham3,spin-2,aspects,Hassan,Hassan2,massive,pauli,covariant,Kluson,Rosen}).  Along these lines, it has long been known that the first massive gravity theory was introduced circa 1939 by Pauli and Fierz in Ref. \cite{pauli}, where they presented a linear action with respect to a spin-2 field on a flat space-time background. The Fierz-Pauli theory describes five degrees of freedom of positive energy in four dimensions at the linear level whereas General Relativity has two degrees of freedom.  However, Boulware and Deser studied some specific fully non-linear massive gravity theories and  pointed out that a general non-linear theory of massive gravity generically contains six propagating degrees of freedom. While the linear theory has five degrees of freedom, the non-linear theories  studied by these authors turned out to have an extra  degree of freedom, which however is unphysical as it has a negative kinetic energy and renders the whole theory unstable: it was therefore called the Boulware-Deser ghost \cite{ghost}. After a great effort,  a non-linear theory free of such a  ghost field was at last obtained by de Rham, Gabadadze and Tolley (dRGT) \cite{Rham,Rham1, Rham3}. The advantage of the  dRGT model is that it contains two dynamical constraints that eliminate both the ghost field and its canonically-conjugate momentum. The absence of the Boulware-Deser ghost was shown explicitly by counting  the degrees of freedom in the framework of the Hamiltonian formalism \cite{Rham2,Hassan,Hassan2,Kluson,Rosen}. Unfortunately, the Hamiltonian analysis of these models remains quite complex and, therefore,  their symmetry properties have not been studied yet via  first-class constraints. On the other hand, in the study of some topics of General Relativity, such as massive gravity,  it is always useful to consider toy models that share the conceptual foundations of the four-dimensional theories, but at the same time are free of  technical difficulties. This is particularly true in three-dimensional (3D) gravity. In this work, we  focus on  the simplest  3D version of a massive gravity theory.

To obtain a realistic 3D-Einstein gravity as compared to the higher-dimensional theory, regarding the local propagating modes, one can modify the theory by adding up  higher-derivative curvature terms in the Einstein-Hilbert (EH) action, which leads to the simplest 3D-massive gravity theory known as Topologically Massive Gravity (TMG). This theory consists of an  EH term, with or without a consmological constant $\Lambda$, plus a  parity-violating gravitational Chern-Simons (CS) term with coefficient $\frac{1}{\mu}$ \cite{deser2,DJG,DJG2,witten}. At the linear level, this theory describes a single massive state of helicity +2 or -2 (depending on the relative sign between the EH and CS terms) in Minkowski background\footnote{ In the presence of a cosmological constant, Minkowski space-time is no longer a vacuum solution and the new maximally symmetric solutions are de Sitter (dS) space-time for positive $\Lambda$ ( dS has isometry group $SO(3,1)$  ) and anti-de Sitter (AdS) space-time for negative $\Lambda$ (AdS has isometry group $SO(2,2)$).} \cite{Strominger}  and defines a unitary irreducible representation of the 3D Poincar\'e group \cite{deser1}.\\

However, while a linearized analysis usually allows a reliable counting of the physical degrees of freedom,  it can yield misleading results in some cases. A Lagrangian/Hamiltonian formulation should provide a way to count the number of local physical degrees of freedom without resorting to linearization,  that is, taking into account all the physical constraints and gauge invariance (i.e. gauge-independence). In this sense,  the identification of the physical degrees of freedom can be addressed by a direct application of Dirac's method for constrained Hamiltonian systems \cite{Dirac}, which systematically separates  all the constraints into first-and second-class ones \cite{teitelboim, Henneaux}. As a consequence, the physical degrees of freedom can be separated from the gauge degrees of freedom, and a generator of the gauge symmetry can be constructed out of a combination of first-class constraints \cite{Castellani}. Furthermore, the bracket structure (Dirac's brackets) to quantize a gauge system  can  be obtained once the second-class constraints are removed. In the case of the massive gravity theories, however, the separation between first-and second-class constraints is a delicate issue, and the system considered in this paper  is not an exception \cite{blagojevic, Mu, Carlip,Grumiller}. In particular, in Ref.  \cite{blagojevic} the Hamiltonian structure of TMG was further analyzed via the Dirac formalism. Indeed, these authors obtain the  secondary first-class-constraint structure of this model with the help of the  theorem: \emph{``If $\phi$ is a first-class constraint, then $\{\phi, H^{T}\}$ is also a first-class constraint''}. Nevertheless, this treatment is quite involved and unsatisfactory.  On the other hand,  the authors of Refs. \cite{Carlip, Grumiller} present a fully Lagrangian analysis, but  the right number of physical degrees of freedom in configuration space can only be obtained once  an ad hoc extra constraint on the basic variables is invoked. This is the main difficulty and it thus worth exploring whether all the necessary constraints can be systematically obtained via a Lagrangian formulation. Thereby, the analysis of the constraints and the gauge symmetry of massive gravity models,  still missing in the literature, is relevant  and  it is thus mandatory to carry out such an analysis to  quantize the theory.

Very interestingly, as an alternative to Dirac's method, Faddeev and Jackiw \cite{Faddeev} proposed a new approach,  which is  geometrically well motivated  and is based on the symplectic structure for constrained systems. This approach, the so-called Faddeev-Jackiw (F-J) symplectic formalism (for a detailed account  see \cite{barcelos,barcelos2,Montani1,Montani2,Pons,modified,non-abelian,hidden}), is useful to  obtain  in an elegant  way several essential elements of a particular  physical theory, such as the physical constraints, the local gauge symmetry, the quantization bracket structure and the number of physical degrees of freedom. It turns out that the F-J approach does not require to classify the constraints into first- and second-class ones. Even more, it   does not invoke Dirac's conjetura. Rather, in this approach,  the quantization brackets can be identified as the elements of the inverse matrix of the symplectic one. For a gauge system, the symplectic matrix remains singular unless a gauge-fixing procedure is introduced. In addition, the generators of the gauge symmetries are given in terms of the zero-modes of the symplectic matrix. In this respect, the F-J symplectic method provides an effective tool for dealing  with gauge theories.

The purpose of this article is to present a detailed F-J  analysis of three-dimensional topologically massive AdS gravity in a completely different context to that presented in Refs. \cite{blagojevic, Mu, Carlip,Grumiller}. In particular, we study the nature of the physical constraints  and  obtain  the gauge symmetry, as well as its generators, under which all the physical quantities must be invariant. Afterwards,  we  obtain both the fundamental quantization brackets and the number of  physical degrees of freedom by introducing an appropriate gauge-fixing procedure.
The remainder of this paper is structured as follows. In section II we briefly review the topologically massive AdS gravity action. Section III is devoted to  explore the nature of the constraints within the Faddeev-Jackiw symplectic framework and  derive the corresponding symplectic matrix.  The  full set of physical constraints of the theory are also obtained. In Section IV, the gauge symmetry and its generators are obtained via the zero-modes of the symplectic matrix. We introduce  gauge-fixing conditions in order to obtain both the quantization bracket structure and the number of physical degrees of freedom in Section V. We conclude with a brief discussion of our results in Section VI.
\section{ Action and equations of motion of Topologically massive gravity }
Our starting point is the action of topologically massive AdS gravity written in the first-order formalism:
\begin{equation}
S[A,e,\lambda]=\int_{\mathcal{M}}\left[2\theta e^{i}\wedge F[A]_{i}-\frac{1}{3}\Lambda f_{ijk}e^{i}\wedge e^{j}\wedge e^{k}+\lambda^{i}\wedge T_{i}+\frac{\theta}{\mu}A^{i}\wedge \left(dA_{i}+\frac{1}{3}f_{ijk}A^{j}\wedge A^{k}\right)\right],
\label{1}
\end{equation}
where $\mu$ is the Chern-Simons parameter, $\theta=1/16\pi G$ with $G$ the  3D Newton's constant, and $\Lambda$ is a cosmological constant such that $\Lambda=-1/l^{2}$, where $l$ is the AdS radius \cite{Strominger}.  Furthermore, the fundamental fields of this action are: the dreibein 1-form $e^{i}=e_{\mu}^{i}dx^{\mu}$ that determines a space-time metric via  $g_{\mu\nu}=e_{\mu}{^{i}}e_{\nu}{^{j}}\eta_{ij}$; the  auxiliary field  1-form $\lambda^{i}$ that ensures that the torsion vanishes $T_{i}=0$ \cite{Carlip2, Carlip3}; and the dualized spin-connection $A^{i}=A_{\mu}{^{i}}dx^{\mu}$ valued on the adjoint representation of the Lie group $SO(2,2)$, so  that,  it admits an invariant totally anti-symmetric tensor $f_{ijk}$. The connection acts  on internal indices and defines a derivative operator:
\begin{equation}
D_{\mu}V^{i}\equiv \partial_{\mu}V^{i}+f^{i}{_{jk}}A_{\mu}{^{j}}V^{k},
\end{equation}
where $\partial$ is a fiducial derivative operator. Finally, $T_{i}$ is the local Lorentz covariant torsion 2-form and $F_{i}$ is the curvature 2-form  of the spin connection $A^{i}$, which explicitly read
\begin{equation}
T_{i}\equiv de_{i}+f{_{ijk}}A^{j}\wedge e^{k},\hphantom{111}F_{i}\equiv dA_{i}+\frac{1}{2}f{_{ijk}}A^{j}\wedge A^{k}.
\end{equation}
The convention adopted is the standard one, that is,  Greek indices  refer to spacetime coordinates and Latin letters correspond to Lorentz indices. The equations of motion that can be extracted by varying  the action (\ref{1}) with respect to $e^{i}$, $A^{i}$ and $\lambda^{i}$, respectively, in addition to some total derivative terms, are given by
\begin{eqnarray}
\left(\delta e\right)^{\alpha i}&=&\epsilon^{\alpha\nu\rho}\left(2\theta F_{\nu\rho}{^{i}}+D_{\nu}\lambda_{\rho}{^{i}}-\Lambda f^{i}{_{jk}}e_{\nu}{^{j}}e_{\rho}{^{k}}\right)=0,\label{mot1}\\
\left(\delta A\right)^{\alpha i}&=&\epsilon^{\alpha\nu\rho}\left(2\theta T_{\nu\rho}{^{i}}+f^{i}{_{jk}}\lambda_{\nu}{^{j}}e_{\rho}{^{k}}+2\theta\mu^{-1}F_{\nu\rho}{^{i}}\right)=0,\label{mot2}\\
\left(\delta \lambda\right)^{\alpha i}&=&\epsilon^{\alpha\nu\rho}T_{\nu\rho}{^{i}}=0.\label{mot3}
\end{eqnarray}
One can note that  Eq. (\ref{mot3}) is the condition for the compatibility of $A_{\mu}{^{i}}$ and $e_{\mu}{^{i}}$, which implies
\begin{equation}
A_{\mu}{^{ij}}=-e^{\nu j}\partial_{\mu}e_{\nu}{^{i}}+\Gamma_{\alpha\mu}^{\beta}e_{\beta}^{i}e^{\alpha j},
\end{equation}
with $\Gamma_{\alpha\mu}^{\beta}$ the Christoffel symbols of the metric $g_{\mu\nu}$, and $A_{\mu}{^{ij}}$ the standard connection obtained by dualizing  the $f$-tensor, $A_{\mu}{^{ij}}=-f^{ij}{_{k}}A_{\mu}{^{k}}$.  Moreover, by inserting  Eq. (\ref{mot3}) into Eq. (\ref{mot2}), one can solve for the Lagrangian multiplier $\lambda_{\mu}{^{i}}$ in terms of the 3D Schouten tensor of the manifold ${\mathcal{M}}$:
\begin{equation}
\lambda_{\mu}{^{i}}=2\theta\mu^{-2}S_{\mu\nu}e^{i\nu}\hphantom{11}\mathrm{with}\hphantom{11} S_{\mu\nu}=R_{\mu\nu}-\frac{1}{4}g_{\mu\nu}R.\label{sol}
\end{equation}
Here we have made use of the fact that the internal and space-time curvature tensors $F_{\mu\nu}{^{ij}}$ and $R_{\mu\nu}{^{\alpha\beta}}$ are related by
\begin{equation}
R^{\alpha\beta}{_{\mu\nu}}=e^{\alpha}{_{i}}e^{\beta}{_{j}}F_{\mu\nu}{^{ij}}\hphantom{11}\mathrm{with}\hphantom{11}F_{\mu\nu}{^{ij}}=-f^{ij}{_{k}}F_{\mu\nu}{^{k}}.
\end{equation}
After plugging these results into  Eq. (\ref{mot1}) and  a lengthy calculation, one can find  the field equation of  TMG \cite{deser2} in the second-order formalism:
\begin{equation}
G_{\mu\nu}+\frac{1}{\mu}C_{\mu\nu}=0,
\end{equation}
where $G_{\mu\nu}$ is the cosmological-constant-modified Einstein tensor defined as
\begin{equation}
G_{\mu\nu}\equiv R_{\mu\nu}-\frac{1}{2}g_{\mu\nu}R+\Lambda g_{\mu\nu},
\end{equation}
and $C_{\mu\nu}$ is the symmetric traceless Cotton tensor given by
\begin{equation}
C_{\mu\nu}\equiv \epsilon_{\mu}{^{\alpha\beta}}\nabla_{\alpha}\left(R_{\beta\nu}-\frac{1}{4}g_{\beta\nu}R\right),
\end{equation}
where $\nabla$ is the covariant derivative defined by $\Gamma$. Considering small perturbations around an anti-de Sitter background, this theory describes the presence of a single massive graviton mode \cite{deser2,Strominger,deser1}.  However, from a theoretical point of view, it is better to checkout the validity of such rough arguments by a careful Hamiltonian or Lagrangian analysis at nonlinear order.
\section{The nature of the constraints in the Faddeev-Jackiw Symplectic framework}
In order to apply the Faddeev-Jackiw's symplectic approach \cite{Faddeev}, throughout this work we  take the spacetime $\mathcal{M}$ to be globally hyperbolic such that it may be foliated as $\mathcal{M}\simeq\Sigma\times \Re$, where $\Sigma$ corresponds to a Cauchy's surface without boundary $(\partial\Sigma = 0)$ and $\Re$ represents an evolution parameter. By performing a $2 + 1$ splitting  of our fields without breaking the internal symmetry,  the TMG action (\ref{1}) acquires the form,
\begin{eqnarray}
S[A,e,\lambda]&=&\int\left[\epsilon^{ab}\theta\left(\frac{1}{\mu}A_{bi}+2e_{bi}\right)\dot{A}^{i}{_{a}} + \epsilon^{ab}\lambda_{ib}\dot{e}^{i}{_{a}} + \epsilon^{ab}e^{i}{_{0}}\left(\theta F_{abi} + D_{a}\lambda_{bi}-\Lambda f_{ijk}e_{a}{^{j}}e_{b}{^{k}}\right) \right.\nonumber \\
&&+\left. \epsilon^{ab}A^{i}{_{0}}\left(\theta T_{abi} + \frac{1}{\mu}\theta F_{abi} +f_{ijk}\lambda^{j}{_{a}}e^{k}{_{b}}\right)+ \frac{1}{2}\epsilon^{ab}\lambda^{i}{_{0}}T_{abi}\right]d^{3}x,\label{2}
\end{eqnarray}
 up to a boundary term.  Here $F_{ab}{^{i}}=\partial_{a}A_{b}^{i}- \partial_{b}A_{a}^{i}+f^{i}{_{jk}}A_{a}{^{j}}A_{b}{^{k}}$ is the field strength of $ A_{a}{^{i}}$, $T_{ab}{^{i}}=D_{a}e_{b}{^{i}}-D_{b}e_{a}{^{i}}$ and  $D_{a}\lambda_{b}{^{i}}=\partial_{a}\lambda_{b}{^{i}}+ f^{i}{_{ij}}A_{a}{^{j}}\lambda_{b}{^{k}}$. Besides $a,b, c, ... $ are space coordinates and the dot denotes a derivative with respect to the evolution parameter. We can read off the Lagrangian density from  (\ref{2}) as
\begin{eqnarray}
\mathcal{L}^{(0)}&=&\epsilon^{ab}\theta\left(\frac{1}{\mu}A_{bi}+2e_{bi}\right)\dot{A}^{i}{_{a}}+ \epsilon^{ab}\lambda_{ib}\dot{e}^{i}{_{a}} + \epsilon^{ab}e^{i}{_{0}}(\theta F_{abi} + D_{a}\lambda_{bi}-\Lambda f_{ijk}e_{a}{^{j}}e_{b}{^{k}}) + \frac{1}{2}\epsilon^{ab}\lambda^{i}{_{0}}T_{abi}\nonumber \\
&&+\epsilon^{ab}A^{i}{_{0}}\left(\theta T_{abi} + \frac{1}{\mu}\theta F_{abi} +f_{ijk}\lambda^{j}{_{a}}e^{k}{_{b}}\right).\label{3}
\end{eqnarray}
In particular, this Lagrangian density can be expressed compactly as
\begin{equation}
\mathcal{L}^{(0)}=a_{I}^{(0)}\dot{\xi}^{(0)I}-V^{(0)},\label{simp_form}
\end{equation}
where  an initial set of symplectic variable is introduced as follows
\begin{equation}
{\xi}{^{(0)I}} = (A^{i}{_{a}}, A^{i}{_{0}}, e^{i}{_{a}},e^{i}{_{0}}, \lambda^{i}{_{a}}, \lambda^{i}{_{0}}),\label{vari}
\end{equation}
 which allows us to identify the corresponding symplectic one-form
\begin{equation}
{a}^{(0)}{_{I}} = (\epsilon^{ab}\theta\left(\frac{1}{\mu}A_{bi}+2e_{bi}\right), 0, \epsilon^{ab}\lambda_{bi}, 0,0, 0),\label{for}
\end{equation}
whereas the symplectic potential reads as
\begin{equation}
V^{(0)}= -\epsilon^{ab}e^{i}{_{0}}(\theta F_{abi} + D_{a}\lambda_{bi}-\Lambda f_{ijk}e_{a}{^{j}}e_{b}{^{k}}) - \frac{1}{2}\epsilon^{ab}\lambda^{i}{_{0}}T_{abi}-\epsilon^{ab}A^{i}{_{0}}\left(\theta T_{abi} + \frac{1}{\mu}\theta F_{abi} +f_{ijk}\lambda^{j}{_{a}}e^{k}{_{b}}\right).
\end{equation}
On the other hand, the corresponding equations of motion arising from the above Lagrangian (\ref{simp_form}) can be written as
\begin{equation}
f_{IJ}^{(0)}\dot{\xi}^{(0)J}-\frac{\delta}{\delta\xi^{(0)I}} V(\xi)^{(0)}=0,\label{eq-mot}
\end{equation}
with $f_{IJ}^{(0)}\equiv\frac{\delta}{\delta\xi^{(0)I}}a^{(0)}_{J}-\frac{\delta}{\delta\xi^{(0)J}}a^{(0)}_{I}$ the  two-form symplectic matrix associated with $\mathcal{L}^{(0)}$, which is clearly antisymmetric.
 By using the symplectic variables (\ref{vari}) and (\ref{for}), we find that the corresponding symplectic matrix $f_{IJ}^{(0)}(x,y)$ can be written as
\begin{eqnarray}
\label{eq}
\left(
  \begin{array}{cccccc}
 2\frac{\theta}{\mu}\epsilon^{ab}\eta_{ij}    &   0   &  -2\theta\epsilon^{ab}\eta_{ij}   &  0     & 0  &\quad  0   \\
 0 &  0   & 0   &  0 &  0   &  0 \\
2\theta\epsilon^{ab}\eta_{ij}   &  0   &   0      &   0   &    -\epsilon^{ab}\eta_{ij}   &  0 \\
    0   & 0   &   0    & 0  & 0 & 0 \\
0  & 0  &    \epsilon^{ab}\eta_{ij}  &  0  &  0 & 0  \\
0   &  0  &  0  &  0   &  0 &  0
 \end{array}
\right)\delta^{2}(x-y).\label{f_0}
\end{eqnarray}
It is not difficult to see that the matrix $f_{IJ}^{(0)}$ is degenerate in the sense  that there are more degrees of freedom in the equations of
motion (\ref{eq-mot}) than physical degrees of freedom in the theory. In this case, there are constraints that must remove the unphysical degrees of freedom. In this formalism the constraints  emerge as algebraic relations necessary to maintain the consistency of the equations of motion. Moreover, it is straightforward to determine that the zero-modes of the  singular matrix (\ref{f_0})  are $(v_{1}^{(0)})^{I}= (0,v^{A{^{i}{_{0}}}}, 0, 0, 0, 0)$, $(v_{2}^{(0)})^{I}=(0, 0, 0, v^{e{^{i}{_{0}}}}, 0, 0)$ and $(v_{3}^{(0)})^{I}=(0, 0, 0, 0, 0, v^{\lambda{^{i}{_{0}}}})$, with non-vanishing arbitrary components $v^{A{^{i}{_{0}}}},v^{e{^{i}{_{0}}}}$ and $v^{\lambda{^{i}{_{0}}}}$, respectively.

The zero-modes satisfy the equation $(v^{(0)}_{1,2,3})^{I}f_{IJ}^{(0)}=0$, therefore from the equation of motion (\ref{eq-mot}), we have the following constraint relations:
\begin{eqnarray}
\int\ dx^{2}(v_{1}^{(0)})^{T}_{J}\frac{\delta}{\delta\xi^{J}}\int\ dy^{2} V^{(0)}  &=& v^{A_{0}^{i}}\left(\theta\epsilon^{ab}T_{abi} + \frac{\theta}{\mu}\epsilon^{ab}F_{abi} + \epsilon^{ab} f_{ijk}\lambda^{j}{_{a}}e^{k}{_{b}}\right)=  0,\\
\int\ dx^{2}(v_{2}^{(0)})^{T}_{J}\frac{\delta}{\delta\xi^{J}}\int\ dy^{2}V^{(0)} &=&v^{e_{0}^{i}}\left( \theta\epsilon^{ab}F_{abi} + \epsilon^{ab}D_{a}\lambda_{bi}-\Lambda \epsilon^{ab}f_{ijk}e_{a}{^{j}}e_{b}{^{k}}\right) = 0, \\
\int\ dx^{2}(v_{3}^{(0)})^{T}_{J}\frac{\delta}{\delta\xi^{J}}\int\ dy^{2}V^{(0)} &=& v^{\lambda_{0}^{i}}\left(\frac{1}{2}\epsilon^{ab} T_{abi}\right) = 0,
\end{eqnarray}
where $v^{A_{0}^{i}}$, $v^{e_{0}^{i}}$ and $v^{\lambda_{0}^{i}}$ are arbitrary functions. The constraints become
\begin{eqnarray}
\Xi_{i}^{(0)}  &=& \theta\epsilon^{ab}T_{abi} + \frac{\theta}{\mu}\epsilon^{ab}F_{abi} + \epsilon^{ab} f_{ijk}\lambda^{j}{_{a}}e^{k}{_{b}}=  0,\label{p1}\\
\Theta_{i}^{(0)} &=& \theta\epsilon^{ab}F_{abi} + \epsilon^{ab}D_{a}\lambda_{bi}-\Lambda \epsilon^{ab}f_{ijk}e_{a}{^{j}}e_{b}{^{k}} = 0,\label{p2} \\
\Sigma_{i}^{(0)} &=&  \frac{1}{2}\epsilon^{ab} T_{abi} = 0.\label{p3}
\end{eqnarray}
Now, according to the methodology of the symplectic framework, we will analyze whether there are new constraints. To achieve this, we demand  stability (consistency condition) of the constraints (\ref{p1}), (\ref{p2}) and (\ref{p3}), which guarantees their time-independence. Since $\Xi_{i}$, $\Theta_{i}$ and $\Sigma_{i}$ depend only on the set of symplectic variables $\xi^{(0)I}$, the  consistency condition  can be written as
\begin{equation}
\dot{\Omega}^{(0)}=\frac{\delta\Omega^{(0)}}{\delta\xi^{(0)I}}\dot{\xi}^{(0)I}=0\hphantom{111}\mathrm{with}\hphantom{111}\Omega^{(0)}=\Xi_{i}^{(0)},\Theta_{i}^{(0)}, \Sigma_{i}^{(0)}.\label{consistency}
\end{equation}
Therefore the consistency of the constraints $\Omega^{(0)}$, together with the equations of motion (\ref{eq-mot}) can be generally rewritten as
\begin{equation}
f_{KJ}^{(1)}\dot{\xi}^{(0)J}=Z^{(1)}_{K}(\xi),
\label{p}
\end{equation}
with
\begin{eqnarray}
f_{KJ}^{(1)}=\left(
\begin{array}{cc}
f^{(0)}_{IJ} \\
\frac{\delta}{\delta\xi^{(0)I}}\Omega^{(0)}
\end{array}\right)\hphantom{111}\mathrm{and}\hphantom{111}Z^{(1)}_{K}=
\left(
\begin{array}{cc}
\frac{\delta }{\delta\xi^{(0)I}}V^{(0)} \\
0\\
0\\
0
\end{array}
\right).
\end{eqnarray}
Furthermore, the new  matrix $f_{KJ}^{(1)}$ can be written as
\begin{eqnarray}
\label{mat1}
&&
\left(
 \begin{array}{cccccc}
 2\frac{\theta}{\mu}\eta_{ij}    &  0   &  -2\theta\eta_{ij}   &0     & 0  &  0   \\
 0 &  0   & 0   &  0 &  0   &   0 \\
2\theta\eta_{ij}   &  0   & 0      &  0   &   -\eta_{ij}   &  0 \\
 0   &  0   &  0    & 0  &  0 & 0 \\
0  &  0  & \eta_{ij}  & 0 & 0 &0 \\
0   &   0  & 0  & 0   &  0 &  0 \\
2\frac{\theta}{\mu}(\eta_{ij}\partial_{a}-f_{ijk}A^{k}{_{a}} -\mu f_{ijk}e^{k}{_{a}}) & 0 &
2\theta(\eta_{ij}\partial_{a}-f_{ijk}A^{k}{_{a}}-\frac{1}{2\theta}f_{ijk}\lambda^{k}{_{a}}) & 0 & -f_{ijk}e^{k}{_{a}}
& 0 \\
2\theta(\eta_{ij}\partial_{a}-f_{ijk}A^{k}{_{a}}-\frac{1}{2\theta}f_{ijk}\lambda^{k}{_{a}}) & 0 & 2\Lambda f_{ijk}e_{a}{^{k}} & 0 & (\eta_{ij}\partial_{a}-f_{ijk}A^{k}{_{a}}) & 0\\
-f_{ijk}e^{k}{_{a}} & 0 & \left(\eta_{ij}\partial_{a} -f_{ijk}A^{k}{_{a}}\right) & 0 & 0 & 0
\end{array}
\right)\nonumber\\
&&\times\epsilon^{ab} \delta^{2}(x-y).
\label{30}
\end{eqnarray}
It is clear that  $f_{KJ}^{(1)}$ is not a square matrix, however, it has linearly independent zero-modes, which turn out to be
\begin{eqnarray}
(v_{1}^{(1)}){^{K}}&=& \left(-\partial_{a}\eta^{j}_{m}-f^{j}{_{lm}}A^{l}{_{a}}, 0, -f^{j}{_{lm}}e^{l}{_{a}}, 0, -f^{j}{_{lm}}\lambda_{a}{^{l}}, 0, \eta^{j}_{m}, 0, 0\right), \label{7.1}\\
(v_{2}^{(1)}){^{K}}&=& \left(-\frac{\mu}{2\theta}f^{j}{_{lm}}\lambda^{l}{_{a}}, 0,-\partial_{a}\eta^{j}_{m} - f^{j}{_{lm}}A^{l}{_{a}}, 0,  f^{j}{_{lm}}\left(\mu\lambda^{l}{_{a}}+2\Lambda e^{l}{_{a}}\right), 0, 0, \eta^{j}_{m}, 0\right),\label{7.2}\\
(v_{3}^{(1)}){^{K}}&=& \left(-\frac{\mu}{2\theta}f^{j}{_{lm}}e^{l}{_{a}}, 0, 0, 0, -\partial_{a}\eta^{j}_{m} - f^{j}{_{lm}}A^{l}{_{a}} +\mu f^{j}{_{lm}}e^{l}{_{a}}, 0, 0, 0, \eta^{j}_{m}\right),
\label{7.3}
\end{eqnarray}
such that $(v^{(1)}_{1,2,3})^{K}f_{KJ}^{(1)}=0$. By using the symplectic potential, we find that  the matrix $Z_{K}^{(1)}$ is given by
\begin{eqnarray}
\label{eq}
\left(
  \begin{array}{c}
-2\theta \left(D_{a}e{_{0j}}+\frac{1}{\mu}D_{a}A{_{0j}}\right)+f{_{jlm}}\left(e_{0}{^{l}}\lambda_{a}{^{m}}+\right(\lambda_{0}{^{l}}+2\theta A_{0}{^{l}}\left)e_{a}{^{m}}\right)\\
\Xi^{(0)}_{i} \\
-D_{a}\lambda{_{0j}}-2\theta D_{a}A{_{0j}} + f{_{jlm}}A^{l}{_{0}}\lambda^{m}{_{a}}-2\Lambda f{_{jlm}}e^{l}{_{0}}e^{m}{_{a}}\\
\Theta^{(0)}_{i} \\
-D_{a}e{_{0j}}+ f{_{jlm}}A^{l}{_{0}}e^{m}{_{a}}\\
\Sigma^{(0)}_{i} \\
0 \\
0 \\
0 \\
\end{array}
\right)\epsilon^{ab} \delta^{2}(x-y).\nonumber\\
\label{z}
\end{eqnarray}
By multiplying  both sides  of Eq.(\ref{p}) by the zero-modes of the matrix $f_{KJ}^{(1)}$, and evaluating at $\Omega^{(0)}=0$, we get the following covariant constraint relations (the integration symbols $\int$ is omitted for clarity):
\begin{eqnarray}
(v^{(1)}_{1})^{K}Z^{(1)}_{K}\mid_{\Omega^{(0)}=0}&=&0,\\
(v^{(1)}_{2})^{K}Z^{(1)}_{K}\mid_{\Omega^{(0)}=0}&=&-\frac{1}{2\theta}\mu\epsilon^{\alpha\beta\gamma}\lambda_{\alpha i}e_{\beta}{^{j}}\lambda_{\gamma j},\label{otra}\\
(v^{(1)}_{3})^{K}Z^{(1)}_{K}\mid_{\Omega^{(0)}=0}&=&\frac{1}{2\theta}\mu\epsilon^{\alpha\beta\gamma}e_{\alpha i}e_{\beta}{^{j}}\lambda_{\gamma j}.
\label{17}
\end{eqnarray}
The substitution   $\Omega^{(0)}=0$ guarantees that these constraints will drop from the remainder of the calculation. Then,  from (\ref{otra}) and (\ref{17}), together with the invertibility of $e_{\alpha i}$  and $\lambda_{\alpha i}$, we finally obtain
\begin{equation}
\Phi^{\alpha}=\epsilon^{\alpha\beta\gamma}e_{\beta}{^{j}}\lambda_{\gamma j}=0,\label{ter}
\end{equation}
which are known as symmetry conditions \cite{spin-2} and play a crucial role in the relation of the metric
and tetrad formulations of massive gravity theories and multi-bigravity ones.  Furthermore, one finds that the equation (\ref{ter}) can be  split  into two equations:
\begin{eqnarray}
\Phi^{a}&=&\epsilon^{ab}\left(e^{i}{_{0}}\lambda_{ib} - e^{i}{_{b}}\lambda_{i0}\right)=0,\label{10}\\
\Phi^{0}&=&\epsilon^{ab}e^{i}{_{a}}\lambda_{ib}=0.\label{26}
\end{eqnarray}
We can see that the Eq. (\ref{10}) has fixed fields $e_{0}^{i}$ and $\lambda_{0}^{i}$, whereas  Eq. (\ref{26})  gives  us one more constraint. This agrees completely with what was found in \cite{blagojevic} by means of the Dirac procedure, however, in that formalism the constraints (\ref{10}) and (\ref{26}) arise as tertiary constraints, whereas in \cite{Carlip,Grumiller} the constraint (\ref{26}) was introduced by hand.  Now,  by imposing  the stability condition on the new constraint (\ref{26}), we have the following equation:
\begin{eqnarray}
f_{KJ}^{(2)}\dot{\xi}^{(0)J}=Z^{(2)}_{K}(\xi),\label{27}
\end{eqnarray}
where the matrices $f_{KJ}^{(2)}$ and $Z^{(1)}_{K}$ can be expressed as
\begin{eqnarray}
f_{KJ}^{(2)}=\left(
\begin{array}{cc}
f^{(1)}_{IJ} \\
\frac{\delta}{\delta\xi^{(0)I}}\Phi^{0}
\end{array}\right)\hphantom{111}\mathrm{and}\hphantom{111}Z^{(2)}_{K}=
\left(
\begin{array}{cc}
Z^{(1)}_{K} \\
0
\end{array}
\right).
\end{eqnarray}
Consequently, the new  matrix $f_{IJ}^{(2)}$ is given by
\begin{eqnarray}
\label{eq}
&&
\left(
 \begin{array}{cccccc}
 2\frac{\theta}{\mu}\eta_{ij}    &  0   &  -2\theta\eta_{ij}   &0     & 0  &  0   \\
 0 &  0   & 0   &  0 &  0   &   0 \\
2\theta\eta_{ij}   &  0   & 0      &  0   &   -\eta_{ij}   &  0 \\
 0   &  0   &  0    & 0  &  0 & 0 \\
0  &  0  & \eta_{ij}  & 0 & 0 &0 \\
0   &   0  & 0  & 0   &  0 &  0 \\
2\frac{\theta}{\mu}(\eta_{ij}\partial_{a}-f_{ijk}A^{k}{_{a}} -\mu f_{ijk}e^{k}{_{a}}) & 0 &
2\theta(\eta_{ij}\partial_{a}-f_{ijk}A^{k}{_{a}}-\frac{1}{2\theta}f_{ijk}\lambda^{k}{_{a}}) & 0 & -f_{ijk}e^{k}{_{a}}
& 0 \\
2\theta(\eta_{ij}\partial_{a}-f_{ijk}A^{k}{_{a}}-\frac{1}{2\theta}f_{ijk}\lambda^{k}{_{a}}) & 0 & 2\Lambda f_{ijk}e_{a}{^{k}} & 0 & (\eta_{ij}\partial_{a}-f_{ijk}A^{k}{_{a}}) & 0\\
-f_{ijk}e^{k}{_{a}} & 0 & \left(\eta_{ij}\partial_{a} -f_{ijk}A^{k}{_{a}}\right) & 0 & 0 & 0\\
0& 0& \lambda_{aj}&0 &e_{aj}&0
\end{array}
\right)\nonumber\\
&&\times\epsilon^{ab} \delta^{2}(x-y).
\label{30}
\end{eqnarray}
One can easily verify  that  $f_{IJ}^{(2)}$ is also a singular matrix that has the following linearly independent zero-modes:
\begin{eqnarray}
(v_{1}^{(2)}){^{J}}{^{}}&=& \left(-\partial_{a}\eta^{j}_{m}-f^{j}{_{lm}}A^{l}{_{a}}, 0, -f^{j}{_{lm}}e^{l}{_{a}}, 0, -f^{j}{_{lm}}\lambda_{a}{^{l}}, 0, \eta^{j}_{m}, 0, 0,0\right),\label{30} \\
(v_{2}^{(2)}){^{J}}{^{}}&=& \left(-\frac{\mu}{2\theta}f^{j}{_{lm}}\lambda^{l}{_{a}}, 0,-\partial_{a}\eta^{j}_{m} - f^{j}{_{lm}}A^{l}{_{a}}, 0,  f^{j}{_{lm}}\left(\mu\lambda^{l}{_{a}}+2\Lambda e^{l}{_{a}}\right), 0, 0, \eta^{j}_{m}, 0,0\right),\label{31}\\
(v_{3}^{(2)}){^{J}}{^{}}&=& \left(-\frac{\mu}{2\theta}f^{j}{_{lm}}e^{l}{_{a}}, 0, 0, 0, -\partial_{a}\eta^{j}_{m} - f^{j}{_{lm}}A^{l}{_{a}} +\mu f^{j}{_{lm}}e^{l}{_{a}}, 0, 0, 0, \eta^{j}_{m},0\right),\label{32}\\
(v_{4}^{(2)}){^{J}}{^{}}&=& \left(\mu e_{a}{^{j}}, 0,e_{a}{^{j}}, 0,  -\left(\lambda_{a}{^{j}}+2\theta\mu e_{a}{^{j}}\right), 0, 0, 0, 0,\eta_{m}^{j}\right).\label{33}
\end{eqnarray}
After performing the contraction of the both of  (\ref{27}) with the  new zero-modes, it is not difficult to see that the zero-modes (\ref{30}), (\ref{31}) and (\ref{32}) do not generate any new constraint, whereas from the zero-mode $(v_{4}^{(2)}){^{J}}$ we have the following constraint relation:
\begin{equation}
(v^{(2)}_{4})^{K}Z^{(2)}_{K}\mid_{\Omega^{(0)},\Omega^{(1)}=0}=\epsilon^{\alpha\beta\nu}f_{ijk}e_{\alpha}^{i}e_{\beta}^{j}\left(\Lambda e_{\nu}^{k}+\mu\lambda_{\nu}^{k}\right)=-2{\bf{e}}\left(3\Lambda+\mu\lambda\right)=0,
\label{177}
\end{equation}
where we have used $\epsilon^{\alpha\beta\nu}e_{\alpha}{^{i}}e_{\beta}{^{j}}e_{\nu}{^{k}}={\bf e }f^{ijk}$ with ${\bf e}=\mathrm{det}\mid e_{\alpha}{^{i}}\mid$ and  $\lambda=e_{\alpha}{^{i}}\lambda^{\alpha}{_{i}}$. Hence, from Eq. (\ref{177}) we can identify the following scalar constraint:
\begin{equation}
\Upsilon=3\Lambda+\mu\lambda=0,\label{quartic}
\end{equation}
which is also in agreement with what was obtained in Ref. \cite{blagojevic} via the Dirac procedure, whereas in \cite{Carlip,Grumiller} such a constraint is missing. Once again, we can introduce the consistency condition on (\ref{quartic}) and explore whether there are further constraints in the theory. To this aim, we study the equation
\begin{equation}
f_{KJ}^{(3)}\dot{\xi}^{(0)J}=Z^{(3)}_{K}(\xi).
\end{equation}
It is easy to verify that even after inserting the above constraint into  the matrix $f_{KJ}^{(3)}$ and calculating its zero-modes, no new constraint is obtained. Hence, there are no further constraints in the theory and thus our procedure to obtain new constraints via the consistency condition is done. With the above results  and the F-J method, we can now  introduce the constraints (\ref{p1}), (\ref{p2}), (\ref{p3}), (\ref{26}) and (\ref{quartic}) into the Lagrangian  density (\ref{3}) by means of the corresponding Lagrangian multipliers in order to construct a new one. So, the new symplectic Lagrangian  can be written  as
\begin{eqnarray}
{\mathcal{L}}&=& \epsilon^{ab}\theta\left(\frac{1}{\mu}A_{bi}+2e_{bi}\right)\dot{A}{^{i}{_{a}}} + \epsilon^{ab}\lambda_{ib}\dot{e}{^{i}{_{a}}}- \Xi_{i}\dot{\beta}^{i} - \Theta_{i}\dot{\alpha}^{i}-\Sigma_{i}\dot{\gamma}^{i}  -\Phi^{0}\dot{\varphi}_{0} -\Upsilon\dot{\varphi}-V,\label{37}
\end{eqnarray}
with  $\dot{\alpha}{^{i}}$, $\dot{\beta}{^{i}}$, $\dot{\gamma}{^{i}}$, $\dot{\varphi_{0}}$ and $\dot{\varphi}$ the Lagrangian multipliers relative to the resulting constraints. Furthermore, one can note that the symplectic potential vanishes on the constraint surface since  it turns out to be a linear combination of constraints reflecting the general covariance of the theory, that is, $V=V^{(0)}\mid_{\Omega^{(0)},\Omega^{(1)},\Phi^{0}, \Upsilon=0}=0$. Moreover, from the Lagrangian density (\ref{37}), the new symplectic variable set is taken as
\begin{equation}
\xi^{I}=\left(A^{i}{_{a}}, \beta^{i}, e^{i}{_{a}},\alpha^{i}, \lambda^{i}{_{a}},\gamma^{i},\varphi^{0},\varphi\right),\label{final-variables}
\end{equation}
whose corresponding canonical 1-form is given by
\begin{equation}
a_{I}= \left(\epsilon^{ab}\theta\left(\frac{1}{\mu}A_{bi}+2e_{bi}\right), -\Xi{_{i}},\epsilon^{ab}\lambda_{bi}, -\Theta_{i}, 0, -\Sigma_{i},-\Phi_{0}, -\Upsilon\right).\label{a-variables}
\end{equation}
We can then use the symplectic variables  (\ref{final-variables}) and (\ref{a-variables}) to construct the corresponding square symplectic matrix $f_{IJ}\equiv\frac{\delta}{\delta\xi^{I}}a_{J}-\frac{\delta}{\delta\xi^{J}}a_{I}$, which turns out to be
\begin{eqnarray}
\label{eq}
{
\left(
 \begin{array}{ccccccccc}
 2\frac{\theta}{\mu}\eta_{ij}    &  -2\frac{\theta}{\mu}\nabla_{aij}  &  -2\theta\eta_{ij}   &-2\theta\triangle_{aij}    &   0&  E_{aij}&0&0\\
 2\frac{\theta}{\mu}\nabla_{aij}   &  0   & 2\theta\triangle_{aij}  &  0 &  -E_{aij}   &   0&0&0 \\
2\theta\eta_{ij}   &  -2\theta\triangle_{aij}   & 0      &  -2\Lambda E_{aij}   &   -\eta_{ij}   &  -D_{aij}^{x}&-\lambda_{aj}&-\mu\epsilon_{ad}\lambda^{d}{_{j}} \\
2\theta\triangle_{aij}   &  0  & 2\Lambda E_{aij}  & 0&D_{aij}^{y} &0&0&0 \\
0&E_{aij}   &   \eta_{ij}  & -D_{aij}^{x}  & 0&0   &  -e_{aj}&-\mu\epsilon_{ad}e^{d}{_{j}}\\
-E_{aij} & 0 &D_{aij}^{y} & 0 & 0& 0&0&0 \\
0 & 0 &\lambda_{aj} & 0 & e_{aj}& 0&0&0 \\
0 & 0 & \mu\epsilon_{ad}\lambda^{d}{_{j}} & 0 & \mu\epsilon_{ad}e^{d}{_{j}} & 0&0&0
\end{array}
\right)}\epsilon^{ab} \delta^{2}(x-y).
\end{eqnarray}
Here, we have defined $\nabla_{aij}=\left(D_{aij}-\mu E_{aij}\right)$ and $\triangle_{aij}=\left(D_{aij}-\frac{1}{2\theta}L_{aij}\right)$ with $D_{aij}=\partial_{a}\eta_{ij}-f_{ijk}A^{k}{_{a}}$, $E_{aij}=f_{ijk}e^{k}{_{a}}$ and $L_{aij}=f_{ijk}\lambda^{k}{_{a}}$ respectively. It is worth noting that  the symplectic matrix $f_{IJ}$  remains singular on the constrained surface, and therefore it still has linearly independent zero-modes. Nevertheless, we have shown that no more constraints can be obtained via the consistency conditions. The non-invertibility of $f_{IJ}$ is then due to a gauge symmetry that must be fixed via additional conditions (gauge conditions) meant to remove the singularity. In this way the quantization-bracket structure can be determined and the procedure can be achieved in terms of the physical degrees of freedom.
\section{Gauge transformations}
It is well-know that the concept of gauge symmetry has played a central role in the development of fundamental theories of physical laws. On the other hand, the need to describe the interactions through relativistic dynamics led us to build a covariant language with a gauge symmetry \cite{utiyama}. We thus proceed towards the discussion of the gauge symmetry in the symplectic framework. It is worth noting that, when all the constraints have been considered and the symplectic matrix still has zero-modes but no new constraint can be obtained,  one is led to conclude that the theory  must have a local gauge symmetry, Therefore the zero-modes act as the generators of the corresponding gauge symmetry `$\delta_{G}$',  that is, the components of the zero-modes give the  transformation properties related to the underlying (gauge) symmetry {\cite{barcelos2, Montani1,Montani2}. The local infinitesimal transformations of the symplectic variables generated by $\left(v\right)^{I}$ can be expressed as
\begin{equation}
\delta_{G}\xi^{I}=\left(v_{A}\right)^{I}\epsilon^{A},\label{44}
\end{equation}
where $\left(v_{A}\right)$ are the independent zero-modes  of the singular symplectic matrix $f_{IJ}$ and $\epsilon^{A}$ are the gauge parameters. For the singular symplectic  matrix (\ref{eq}), these  zero-modes turn out to be
\begin{eqnarray}
\left(v_{1}\right)^{I}&=&\left(-\partial_{a}\eta^{j}{_{k}}-f^{j}{_{lk}}A_{a}{^{l}}, \eta^{j}{_{k}},-f^{j}{_{lk}}e_{a}^{l},0,-f^{j}{_{lk}}\lambda_{a}{^{l}},0,0,0,0\right),\label{v1}\\
\left(v_{2}\right)^{I}&=&\left(-\frac{\mu}{2\theta}f^{j}{_{lk}}\lambda_{a}{^{l}}, 0,-\partial_{a}\eta^{j}{_{k}}-f^{j}{_{lk}}A_{a}{^{l}},\eta^{j}{_{k}}, f^{j}{_{lk}}\left(\mu\lambda_{a}{^{l}}+2\Lambda e_{a}{^{l}}\right),0,0,0,0\right),\label{v2}\\
\left(v_{3}\right)^{I}&=&\left(-\frac{\mu}{2\theta}f^{j}{_{lk}}e_{a}{^{l}}, 0, 0, 0,-\partial_{a}\eta^{j}{_{k}}-f^{j}{_{lk}}A_{a}{^{l}}+\mu f^{j}{_{lk}}e_{a}{^{l}},\eta^{j}{_{k}},0,0,0\right),\label{v3}
\end{eqnarray}
which are orthogonal to the gradient of the symplectic potential and at the same time  generate local displacements on the isopotential surface. As one can infer from (\ref{44}), the infinitesimal gauge transformations that leave the original Lagrangian invariant are given by
\begin{eqnarray}
\delta_{G} A_{\alpha}{^{i}}(x)&=&-D_{\alpha}\zeta^{i}-\frac{\mu}{2\theta}f^{i}{_{jk}}\left(e_{\alpha}{^{j}}\varsigma^{k}+\lambda_{\alpha}{^{j}}\kappa^{k}\right),\label{g1}\\
\delta_{G} e_{\alpha}{^{i}}(x)&=&-D_{\alpha}\kappa^{i}-f^{i}{_{jk}}e_{a}^{j}\zeta^{k},\label{g2}\\
\delta_{G} \lambda_{\alpha}{^{i}}(x)&=&-D_{\alpha}\varsigma^{i}-f^{i}{_{jk}}\lambda_{\alpha}{^{j}}\zeta^{k}+\mu f^{i}{_{jk}}\left(\lambda_{a}{^{j}}\kappa^{k}+e_{a}{^{j}}\varsigma^{k}\right)+2\Lambda f^{i}{_{jk}}e_{a}{^{j}}\kappa^{k}\label{g3},
\end{eqnarray}
where $\zeta^{i}$, $\kappa^{i}$ and $\varsigma^{i}$ are the  time-dependent gauge parameters. It is worth remarking that (\ref{g1}), (\ref{g2}) and (\ref{g3}) correspond to the fundamental gauge symmetry of the theory, though the diffeomorphisms have not been found yet. However, it is well-known that an appropriate choice of the gauge parameters does generate the diffeomorphism (on-shell) \cite{utiyama,kibble, blagojevic2}. Let us redefine the gauge parameters as
\begin{equation}
\zeta^{i}=-A^{i}{_{\mu}}\varepsilon^{\mu}, \hphantom{111}\kappa^{i}=-e^{i}{_{\mu}}\varepsilon^{\mu},\hphantom{111}\varsigma^{i}=-\lambda^{i}{_{\mu}}\varepsilon^{\mu},\label{lie}
\end{equation}
with $\varepsilon^{\mu}$  an arbitrary three-vector. Hence,  from the fundamental gauge symmetry (\ref{g3}) and the mapping (\ref{lie}), we obtain
\begin{eqnarray}
\delta_{G}A_{\alpha}{^{i}} &=&\mathfrak{L}_{\varepsilon}A_{\alpha}{^{i}}+\mu\varepsilon^{\mu}\epsilon_{\alpha\mu\nu}\left[\frac{1}{2\theta}\left(\delta A\right)^{\nu i}+\left(\delta \lambda\right)^{\nu i}\right],\nonumber\\
\delta_{G}e_{\alpha}{^{i}} &=&\mathfrak{L}_{\varepsilon}e_{\alpha}{^{i}}-\varepsilon^{\mu}\epsilon_{\alpha\mu\nu}\left(\delta \lambda\right)^{\nu i},\nonumber\\
\delta_{G}\lambda_{\alpha}{^{i}} &=&\mathfrak{L}_{\varepsilon}\lambda_{\alpha}{^{i}}+2\mu\theta\varepsilon^{\mu}\epsilon_{\alpha\mu\nu}\left[\frac{1}{2\mu\theta}\left(\delta e \right)^{\nu i}-\frac{1}{2\theta}\left(\delta A\right)^{\nu i}+\left(\delta \lambda\right)^{\nu i}\right].
\end{eqnarray}
which are precisely (on-shell) diffeomorphisms. In addition, TMG (\ref{1}) is also made  invariant under Poincar\'e  transformations  by construction \cite{kibble, utiyama}. Thus, in order to recover the Poincar\'e  symmetry, we need to map the arbitrary gauge parameters of the fundamental gauge symmetry `$\delta_{G}$' (\ref{g3}) into those of the  Poincar\'e symmetry. This is achieved by a mapping of the gauge parameters \cite{blagojevic2,banerjee,kibble}, e.g.:
\begin{equation}
\zeta^{i}=A^{i}{_{\mu}}\varepsilon^{\mu}+\omega^{i}, \hphantom{111}\kappa^{i}=e^{i}{_{\mu}}\varepsilon^{\mu},\hphantom{111}\varsigma^{i}=\lambda^{i}{_{\mu}}\varepsilon^{\mu}
\end{equation}
such that $\varepsilon^{\mu}$ and $\omega^{i}$ are related to local coordinate translations and local Lorentz rotations, respectively, which together constitute the $6$ independent gauge parameters of Poincar\'e symmetries in 3D. By using this map,  the gauge symmetries  reproduce the Poincar\'e symmetries modulo terms proportional to the equations of motion
\begin{eqnarray}
\delta_{G}e_{\alpha}{^{i}} &=&-\varepsilon^{\mu}\partial_{\mu}e_{\alpha}{^{i}}-e_{\mu}{^{i}}\partial_{\alpha}\varepsilon^{\mu}-f^{i}{_{jk}}e_{\alpha}{^{j}}\omega^{k}+\varepsilon^{\gamma}\epsilon_{\alpha\gamma\nu}\left(\delta \lambda\right)^{\nu i},\nonumber\\
\delta_{G}A_{\alpha}{^{i}} &=&-\partial_{\alpha}\omega^{i}-f^{i}{_{jk}}A_{\alpha}{^{j}}\omega^{k}-\varepsilon^{\mu}\partial_{\mu}A_{\alpha}{^{i}}-A_{\mu}{^{i}}\partial_{\alpha}\varepsilon^{\mu}-\mu\varepsilon^{\gamma}\epsilon_{\alpha\gamma\nu}\left[\frac{1}{2\theta}\left(\delta A\right)^{\nu i}+\left(\delta \lambda\right)^{\nu i}\right],\nonumber\\
\delta_{G}\lambda_{\alpha}{^{i}} &=&-\varepsilon^{\mu}\partial_{\mu}\lambda_{\alpha}{^{i}}-\lambda_{\mu}{^{i}}\partial_{\alpha}\varepsilon^{\mu}-f^{i}{_{jk}}\lambda_{\alpha}{^{j}}\omega^{k}-2\mu\theta\varepsilon^{\gamma}\epsilon_{\alpha\gamma\nu}\left[\frac{1}{2\mu\theta}\left(\delta e \right)^{\nu i}-\frac{1}{2\theta}\left(\delta A\right)^{\nu i}+\left(\delta \lambda\right)^{\nu i}\right],\label{PGT}
\end{eqnarray}
where the equations of motion $\left(\delta e\right)^{\nu i}$, $\left(\delta A\right)^{\nu i}$ and $\left(\delta\lambda\right)^{\nu i}$ are defined in (\ref{mot1})-(\ref{mot3}). We thus conclude that the Poincar\'e symmetry (\ref{PGT}) as well as the diffeomorphisms (\ref{lie}) are not independent symmetries: they are contained indeed in the fundamental gauge symmetry (\ref{g3}) as on-shell symmetries, that is, only when the equations of motion are imposed.  In addition, the generators of such gauge transformations can be represented in terms of the zero-modes, thereby making evident that the zero-modes of the symplectic two-form encode all the information about the gauge structure of this theory.
\section{ The Faddeev-Jackiw brackets and degree of freedom count }
As was already mentioned in Sec. III, in theories with a gauge symmetry, the symplectic matrix obtained at the end of the procedure is still singular. Nevertherless, in order to obtain a non-singular symplectic matrix and to determine the quantization bracket (F-J brackets) structure between the dynamical fields, we must   impose a gauge-fixing procedure, that is, new gauge constraints.  In this case, we now partially fix the gauge by imposing the time-gauge, namely, $A^{i}{_{0}}=0$, $e^{i}{_{0}}=0$, $\lambda^{i}{_{0}}=0 $ and $\varphi_{0}=$ cte (i.e. $\dot{\varphi_{0}}=0$). In this manner, we also introduce new
Lagrange multipliers that enforce these gauge conditions, namely, $\dot{\rho}_{i}$, $\dot{\omega}_{i}$, $\dot{\tau_{i}}$ and $\dot{\sigma}^{0}$.  Thus, the final symplectic Lagrangian after gauge fixing can be written as
\begin{equation}
{\mathcal{L}}=\epsilon^{ab}\theta\left(\frac{1}{\mu}A_{bi}+2e_{bi}\right)\dot{A}{^{i}{_{a}}} + \epsilon^{ab}\lambda_{ib}\dot{e}{^{i}{_{a}}}- \left(\Xi_{i}-\rho_{i}\right)\dot{\beta}^{i} - \left(\Theta_{i}-\omega_{i}\right)\dot{\alpha}^{i}-\left(\Sigma_{i}-\tau_{i}\right)\dot{\gamma}^{i}  -\left(\Phi^{0}-\sigma^{0}\right)\dot{\varphi}_{0}.\label{L.final}
\end{equation}
From the Lagrangian density (\ref{L.final}) one may read off the final set of symplectic variables
\begin{equation}
\xi^{I}=\left(A^{i}{_{a}}, \beta^{i}, e^{i}{_{a}},\alpha^{i}, \lambda^{i}{_{a}},\gamma^{i},\varphi^{0},\rho^{i},\omega^{i},\tau^{i},\sigma_{0}\right),\label{final}
\end{equation}
so that, the corresponding symplectic 1-form is given by
\begin{equation}
a^{}_{I}= \left(\epsilon^{ab}\theta\left(\frac{1}{\mu}A_{bi}+2e_{bi}\right), -\Xi{_{i}}+\rho_{i}, \epsilon^{ab}\lambda_{bi}, -\Theta_{i}+\omega_{i}, 0, -\Sigma_{i}+\tau_{i},-\Phi_{0}+\sigma_{0},0,0,0,0\right).
\end{equation}
After some algebra, we obtain the explicit form of the symplectic matrix $f_{IJ}$
\begin{eqnarray}
&&
{\small{}
\left(
\begin{array}{ccccccccccc}
2\frac{\theta}{\mu}\eta_{ij} & -2\frac{\theta}{\mu}\nabla_{aij} & -2\theta\eta_{ij} &-2\theta\triangle_{aij} & 0& E_{aij}&0&0&0&0&0\\
2\frac{\theta}{\mu}\nabla_{aij} & 0 & 2\theta\triangle_{aij} & 0 & -E_{aij} & 0&0&-\frac{1}{2}\epsilon_{ab}\eta_{ij}&0&0 &0\\
2\theta\eta_{ij} & -2\theta\triangle_{aij} & 0 & -2\Lambda E_{aij} & -\eta_{ij} & -D_{aij}^{x}&-\lambda_{aj}&0&0&0&0 \\
2\theta\triangle_{aij} & 0 & 2\Lambda E_{aij} & 0&D_{aij}^{y} &0&0&0 &-\frac{1}{2}\epsilon_{ab}\eta_{ij}&0&0 \\
0&E_{aij} & \eta_{ij} & -D_{aij}^{x} & 0&0 & -e_{aj}&0&0&0&0\\
-E_{aij} & 0 &D_{aij}^{y} & 0 & 0& 0&0&0&0&-\frac{1}{2}\epsilon_{ab}\eta_{ij}&0 \\
0 & 0 &\lambda_{aj} & 0 & e_{aj}& 0&0&0&0&0&-\frac{1}{2}\epsilon_{ab} \\
0 & \frac{1}{2}\epsilon_{ab}\eta_{ij} & 0 & 0 & 0 & 0&0&0&0&0&0\\
0 & 0 & 0 & \frac{1}{2}\epsilon_{ab}\eta_{ij} & 0 & 0&0&0&0&0&0\\
0 & 0 & 0 &0 & 0 & \frac{1}{2}\epsilon_{ab}\eta_{ij}&0&0&0&0&0\\
0 & 0 & 0 &0 & 0 & 0&\frac{1}{2}\epsilon_{ab}&0&0&0&0
\end{array}
\right)}\nonumber\\
&&\times\epsilon^{ab} \delta^{2}(x-y). \label{regular-matrix}
\end{eqnarray}
It is clear   that such a matrix is not singular. The corresponding  inverse matrix ${f_{IJ}}^{-1}$ is given by
\begin{eqnarray}
&&
{\small{}
\left(
\begin{array}{cccccccccccc}
\frac{\mu}{2\theta}\eta_{ij}&0&0&0&-\mu\eta_{ij}&0&0&D_{aij}^{x}&-\frac{\mu}{2\theta} L_{aij}&-\frac{\mu}{2\theta}E_{aij}&{\mu} e_{a}{_{i}}\\
0&0&0&0&0&0&0&\eta{_{ij}}&0&0&0\\
0&0&0&0&\eta_{ij}&0&0&-E_{aij}&D_{aij}&0 &e_{a}{_{i}}\\
0&0&0&0&0&0&0&0 &\eta_{ij}& 0 & 0\\
\mu\eta_{ij}&0&-\eta_{ij}&0&-2\theta\mu \eta_{ij}&0&0&L_{aij}&-\Diamond_{aij} & -\nabla_{aij} & -\lambda_{ai}\\
0&0&0&0&0&0&0 & 0 & 0 & \eta_{ij} & 0 \\
0&0&0&0&0&0&0 & 0 & 0 & 0 & 1 \\
-D_{aij}^{y}&-\eta_{ij}&E_{aij}&0&-L_{aij}&0&0&0 & 0 & 0 & 0 \\
\frac{\mu}{2\theta}L_{aij}&0&-D_{aij}^{y}&-\eta_{ij}&\Diamond_{aij}&0 & 0 & 0 & -\frac{\mu}{2\theta}L_{aik}L_{b}{^{k}}{_{j}} & 0&0 \\
\frac{\mu}{2\theta}E_{aij}&0&0&0&\nabla_{aij}&-\eta_{ij}&0 & 0 & 0 & -\frac{\mu}{2\theta}E_{aik}E_{b}{^{k}}{_{j}} & e^{aj}\nabla_{aij}& \\
-\mu e_{a}{_{i}}&0&-e_{a}{_{i}}&0&\lambda_{ai}&0&-1 & 0 & 0 & -e^{aj}\nabla_{aij} & 0\\
\end{array}
\right)}\nonumber\\
&&\times\epsilon^{ab} \delta^{2}(x-y),
\label{inv}
\end{eqnarray}
 with $\Diamond_{aij}=\left(\mu L_{aij}+2\Lambda E_{aij}\right)$. In this way, the quantization bracket, dubbed generalized Faddeev-Jackiw bracket, $\{,\}_{F-J}$ between two elements of the symplectic variable set (\ref{final}),  is defined as
\begin{equation}
\{\xi_{I}(x),\xi_{J}(y)\}_{F-J}\equiv\left(f_{IJ}\right)^{-1}.\label{F-J}
\end{equation}
The non-vanishing Faddeev-Jackiw brackets for topologically massive  AdS gravity can now be easily extracted using (\ref{inv}) and (\ref{F-J}). We thus have
\begin{eqnarray}
\{A^{i}{_{a}}(x), A^{j}{_{b}}(y)\}_{F-J} &=& \frac{\mu}{2\theta}\eta^{ij}\delta^{2}(x-y),\label{b-f-j1} \\
\{A^{i}{_{a}}(x), \lambda^{j}{_{b}}(y)\}_{F-J} &=&- \mu\epsilon_{ab}\eta^{ij}\delta^{2}(x-y),\label{b-f-j2} \\
\{\lambda^{i}{_{a}}(x), \lambda^{j}{_{b}}(y)\}_{F-J} &=& 2\theta\mu\epsilon_{ab}\eta^{ij}\delta^{2}(x-y),\label{b-f-j3} \\
\{e^{i}{_{a}}(x), \lambda^{j}{_{b}}(y)\}_{F-J} &=& \epsilon_{ab}\eta^{ij}\delta^{2}(x-y).\label{b-f-j4}
\end{eqnarray}
These F-J brackets correspond to the Dirac brackets reported in \cite{blagojevic}. The canonical quantization $ \left(\{\xi_{I},\xi_{J}\}_{F-J}\rightarrow\frac{1}{i\hbar}\left[\hat{\xi}_{I},\hat{\xi}_{J}\right]\right) $ can be carried out by using the aforementioned brackets given by (\ref{b-f-j1})-(\ref{b-f-j4}). In addition,  we are now ready to perform the counting  of physical  degrees of freedom: starting with $ 18  $ canonical variables $(e^{i}{_{a}}, \lambda^{i}{_{a}}, A^{i}{_{a}})$, we end up with $17$ independent constraints $(\Xi_{i}^{(0)},\Theta_{i}^{(0)}, \Sigma_{i}^{(0)}, \Phi^{0}, e^{i}{_{0}}=0, A^{i}{_{0}}=0, \varphi_{0}=\mathrm{cte})$ after imposing the gauge-fixing term. Therefore, the number of physical  degrees of freedom per space point for 3D Topologically Massive AdS Gravity is one, independently of the value of $\mu$, as it was also found  in \cite{Carlip,Grumiller}.

\section{ Conclusions and discussions}
In the present paper,  the nature of the constraints and gauge structure of the  topologically massive AdS gravity theory was studied from the perspective of the Faddeev-Jackiw symplectic approach. The whole set of independent physical constraints   was identified through the consistency condition and the zero-modes. It was shown that even when all the physical constraints are found, but the symplectic matrix still has zero-modes, that is, when  the zero-modes  are orthogonal  to the gradient of the symplectic potential on the surface of the constraints,  one is led to deduce that the theory  has a local gauge symmetry. Therefore, the zero-modes   straightforwardly  generate the local gauge  symmetry under which all physical quantities are invariant. By mapping the gauge parameters appropriately we have also obtained the Poincar\'e transformations  and the diffeomorphism symmetry. Additionally, we have shown that the time-gauge fixing of the density Lagrangian renders the non-degenerate symplectic matrix $f_{IJ}$. We  then have  identified the quantizaion bracket (F-J brackets) structure and have proved that there is one physical degree of freedom. It is worth remarking that all the results presented here can be applied to the study of  the physical content of models such as massive gravity and bigravity theories  in 2+1 dimensions, in which secondary, tertiary, or higher-order constraints are present.  Such problems are under study and will be published elsewhere   \cite{omar}.  Another line for further research is the  application of the procedure used here to explore conceptual and technical issues of gravity models  in 3+1 dimensions.
\newline
\newline
\newline
\noindent \textbf{Acknowledgements}\\[1ex]
This work has been partially supported by CONACyT under grand number CB-2014-01/240781. We would like to thank G. Tavares-Velasco for reading a draft version of this paper and alerting us to various typos.
\appendix
\section{Faddeev-Jackiw symplectic approach}
In this appendix, we  summarize the main aspects of the Faddeev-Jackiw symplectic approach \cite{Faddeev}, which is based on a  first-order Lagrangian in time derivative. However, this is not a serious restriction because even if the original Lagrangian is not of first-order, it is always possible to introduce variables of auxiliary fields to obtain a first-order one (usually, the canonical momenta are chosen as auxiliary fields). After introducing variables of the auxiliary fields, we can construct a first-order Lagrangian for a physical system  as follows:
\begin{equation}
{\mathcal{L}}(\xi)=a_{I}(\xi)\dot{\xi}^{I}-V(\xi)\hphantom{1111}(I=1,2,3,...,N),\label{f1}
\end{equation}
where $\xi^{I}$ is the so-called symplectic variable, which consists of a combination of the original variables along with some auxiliary fields and the canonical momenta. The term $V(\xi)$, which is called symplectic potential,  is assumed to be free of time derivatives of  $\xi^{I}$, and it is   easy to see that it is the negative of the canonical Hamiltonian. Finally, the function $a_{I}(\xi)$ is the canonical one-form and is the main focus of interest.
The Euler-Lagrange equations of motion for Lagrangian (\ref{f1}) can be written as
\begin{equation}
f_{IJ}\dot{\xi}^{J}-\frac{\partial }{\partial\xi_{I}}V(\xi)=0,\label{f3}
\end{equation}
where $f_{IJ}$ is the so-called symplectic matrix with the following explicit form:
\begin{equation}
f_{IJ}\equiv \frac{\partial}{\partial\xi_{I}}a^{J}-\frac{\partial}{\partial\xi_{J}}a^{I}.\label{f4}
\end{equation}
When this matrix is non-singular,  it can be inverted, and therefore all the symplectic variables can be solved from (\ref{f3})
\begin{equation}
\dot{\xi}_{I}=\left (f\right)_{IJ}^{-1}\frac{\partial}{\partial\xi_{J}}V(\xi).\label{5}
\end{equation}
Otherwise, there are some constraints in the theory. In the method of Faddeev-Jackiw, the above equation can be written as
\begin{equation}
\dot{\xi}_{I}=\{\xi_{I},\xi_{J}\}_{F-J}\frac{\partial V}{\partial\xi_{J}}.\label{paren}
\end{equation}
where the Faddeev-Jackiw bracket $\{,\}_{F-J}$ is defined by
\begin{equation}
\{\xi_{I},\xi_{J}\}_{F-J}=\left(f\right)^{-1}_{IJ}\label{paren-f-j}.
\end{equation}
However, in gauge invariant theories, where in addition to the true dynamical degrees of freedom there are also gauge degrees of freedom,  the symplectic matrix turns out to be singular, which implies that the system is endowed with constraints. In this case,  the matrix $f_{IJ}$ necessarily has some zero-modes  $(v_{k})$ (with $k$ all the linearly independent zero-modes that are found for $f_{IJ}$), where each $(v_{k})$ is a column vector with $N$ entries $(v_{k})^{I}$. By definition, the zero-modes satisfy the following equation
\begin{eqnarray}
\left(v_{k}\right)^{I}f_{IJ}=0,\hphantom{11111}(k=1,2,3,...,\leq N).
\end{eqnarray}
Consequently, the constraints associated with the symplectic matrix are given by
\begin{eqnarray}
\phi_{k}\equiv\left(v{_{k}}\right)^{I}\frac{\partial}{\partial\xi^{I}}V(\xi)=0,\label{p1}
\end{eqnarray}
which shows that the zero-modes of $f_{IJ}$ encode the information of the constraints. Following the prescription of the symplectic formalism, we will analyze whether there are new constraints. To this aim,
we impose a consistency condition on the constraints as
in the Dirac approach:
 \begin{equation}
\dot{\phi}_{k}=\frac{\partial \phi_{k}}{\partial \xi^{I}}\dot{\xi}^{I}=0.\label{f9}
\end{equation}
The consistency condition on the constraints (\ref{f9}) and equations of motion (\ref{f3}) can be rewritten as
\begin{eqnarray}
f^{(1)}_{KJ}\dot{\xi}^{J}= Z^{(1)}_{K}(\xi),\label{11}
\end{eqnarray}
where
\begin{equation}
f^{(1)}_{KJ}=
\left(
\begin{array}{cc}
f_{IJ} \\
\frac{\partial\phi_{k}}{\partial\xi^{J}}
\end{array}
\right)\hphantom{111}{\rm and}\hphantom{111}Z^{(1)}_{K}(\xi)=
\left(
\begin{array}{ccc}
\frac{\partial V}{\partial \xi^{I}} \\
0 \\
\end{array}\label{equation2}
\right),
\end{equation}
The new matrix $f_{IJ}^{(1)}$ is not a square matrix anymore, however, it still contains linearly independent zero-modes $(v_{l}^{(1)})$, which are different from the original ones. Multiplying both sides of Eq. (\ref{11}) by these modes, we get the following constraint relations
\begin{eqnarray}
\left(v{^{(1)}}_{l}\right)^{K}Z^{(1)}_{K}|_{\phi_{k}=0}=0.\label{tt2}
\end{eqnarray}
The substitution $\phi_{k}=0$ guarantees that these constraints
will drop from the remainder of the calculation.  If Eqs. (\ref{tt2}) turn out to fulfill the identity $0 = 0$, then there are no further constraints; otherwise,  the constraints arising from Eq. (\ref{tt2}) are given by
\begin{eqnarray}
\phi_{l}^{(1)}\equiv\left(v{^{(1)}}_{l}\right)^{K}Z^{(1)}_{K}|_{\phi_{k}}=0.\label{t2}
\end{eqnarray}
These new constraints can be treated in the same way as  $\phi_{k}$. In other words, we can now introduce the consistency condition for $\phi_{l}^{(1)}$, as
\begin{equation}
\dot{\phi}_{l}^{(1)}=\frac{\partial \phi_{l}^{(1)}}{\partial \xi^{I}}\dot{\xi^{I}}=0.\label{consis}
\end{equation}
and combine it with Eq. (\ref{11}) in order to construct a set of new
linear equations, from which we explore whether there are more constraints. These steps are repeated until there are no further constraints in the system and  the identities $0 = 0$ are fulfilled.

Once $m$ constraints are obtained after $h$ steps through the consistency conditions of the constraints, we can modify our original Lagrangian (\ref{f1}) by introducing the whole set of constraints multiplied by the corresponding Lagrangian multipliers $\dot{\eta}^{m}$ as follows:
\begin{equation}
{\mathcal{L}}^{(E)}=a_{I}(\xi){}\dot{\xi}^{I}+\phi_{m}(\xi)\dot{\eta}^{m}-V(\xi)^{(E)},\label{alter}
\end{equation}
where  $V(\xi)^{(E)}=V(\xi) |_{\phi_{m}=0}$. We can now also calculate the new symplectic matrix associated with the modified Lagrangian, ${\scriptsize{} f_{IJ}^{(E)}=\partial a_{J}^{(E)}/\partial\xi^{(E)I}-\partial a_{I}^{(E)}/\partial\xi^{(E)J}}$ with ${\scriptsize{} \xi^{(E)I}=(\xi^{I},\eta^{l})}$; this new matrix can be either singular or non-singular.  In the latter case it has an inverse and therefore all the new symplectic variables can be solved as in (\ref{paren}). On the other hand,  for gauge systems, this symplectic matrix is still singular and has no inverse unless some gauge-fixing terms (gauge conditions) are introduced. In this  way, the procedure can be finished and the Faddeev-Jackiw brackets can be identified as in    (\ref{paren-f-j}).

\end{document}